\documentclass[fleqn,twoside,twocolumn,nofootinbib,showkeys]{revtex4-1} 

\usepackage[utf8]{inputenc}
\usepackage{amsmath}
\usepackage{amssymb}
\usepackage{dcolumn}
\usepackage{url}
\usepackage{hyperref}
\usepackage{verbatim}
\usepackage{listings}
\usepackage{fancyvrb}
\usepackage[breakable]{tcolorbox}
\usepackage{spverbatim}
\usepackage{xcolor}
\usepackage{lipsum}
\usepackage{float}
\usepackage{graphicx} 

\definecolor{cellbackground}{HTML}{F7F7F7}
\definecolor{framec}{HTML}{8d81d1}
\definecolor{PY_green}{RGB}{24, 110, 26}
\definecolor{PY_blue}{RGB}{20,20,140}
\definecolor{PY_red}{RGB}{120,20,10}

\makeatletter
        \newbox\Wrappedcontinuationbox 
        \newbox\Wrappedvisiblespacebox 
         
        \newcommand*\Wrappedcontinuationsymbol {{\llap{\tiny$\m@th\hookrightarrow$}}} 
        \newcommand*\Wrappedcontinuationindent {3ex } 
        \newcommand*\Wrappedafterbreak {\kern\Wrappedcontinuationindent\copy\Wrappedcontinuationbox} 
        \newcommand*\Wrappedbreaksatspecials {%
            \def\PYGZus{\discretionary{\char`\_}{\Wrappedafterbreak}{\char`\_}}%
            \def\PYGZob{\discretionary{}{\Wrappedafterbreak\char`\{}{\char`\{}}%
            \def\PYGZcb{\discretionary{\char`\}}{\Wrappedafterbreak}{\char`\}}}%
            \def\PYGZca{\discretionary{\char`\^}{\Wrappedafterbreak}{\char`\^}}%
            \def\PYGZam{\discretionary{\char`\&}{\Wrappedafterbreak}{\char`\&}}%
            \def\PYGZlt{\discretionary{}{\Wrappedafterbreak\char`\<}{\char`\<}}%
            \def\PYGZgt{\discretionary{\char`\>}{\Wrappedafterbreak}{\char`\>}}%
            \def\PYGZsh{\discretionary{}{\Wrappedafterbreak\char`\#}{\char`\#}}%
            \def\PYGZpc{\discretionary{}{\Wrappedafterbreak\char`\%}{\char`\%}}%
            \def\PYGZdl{\discretionary{}{\Wrappedafterbreak\char`\$}{\char`\$}}%
            \def\PYGZhy{\discretionary{\char`\-}{\Wrappedafterbreak}{\char`\-}}%
            \def\PYGZsq{\discretionary{}{\Wrappedafterbreak\textquotesingle}{\textquotesingle}}%
            \def\PYGZdq{\discretionary{}{\Wrappedafterbreak\char`\"}{\char`\"}}%
            \def\PYGZti{\discretionary{\char`\~}{\Wrappedafterbreak}{\char`\~}}%
        } 
        \newcommand*\Wrappedbreaksatpunct {%
            \lccode`\~`\.\lowercase{\def~}{\discretionary{\hbox{\char`\.}}{\Wrappedafterbreak}{\hbox{\char`\.}}}%
            \lccode`\~`\,\lowercase{\def~}{\discretionary{\hbox{\char`\,}}{\Wrappedafterbreak}{\hbox{\char`\,}}}%
            \lccode`\~`\;\lowercase{\def~}{\discretionary{\hbox{\char`\;}}{\Wrappedafterbreak}{\hbox{\char`\;}}}%
            \lccode`\~`\:\lowercase{\def~}{\discretionary{\hbox{\char`\:}}{\Wrappedafterbreak}{\hbox{\char`\:}}}%
            \lccode`\~`\?\lowercase{\def~}{\discretionary{\hbox{\char`\?}}{\Wrappedafterbreak}{\hbox{\char`\?}}}%
            \lccode`\~`\!\lowercase{\def~}{\discretionary{\hbox{\char`\!}}{\Wrappedafterbreak}{\hbox{\char`\!}}}%
            \lccode`\~`\/\lowercase{\def~}{\discretionary{\hbox{\char`\/}}{\Wrappedafterbreak}{\hbox{\char`\/}}}%
            \catcode`\.\active
            \catcode`\,\active 
            \catcode`\;\active
            \catcode`\:\active
            \catcode`\?\active
            \catcode`\!\active
            \catcode`\/\active 
            \lccode`\~`\~ 	
        }
    \makeatother

\let\OriginalVerbatim=\Verbatim
    \makeatletter
    \renewcommand{\Verbatim}[1][1]{%
        \sbox\Wrappedcontinuationbox {\Wrappedcontinuationsymbol}%
        \def\FancyVerbFormatLine ##1{\hsize\linewidth
            \vtop{\raggedright\hyphenpenalty\z@\exhyphenpenalty\z@
                \doublehyphendemerits\z@\finalhyphendemerits\z@
                \strut ##1\strut}%
        }%
        \def\FV@Space {%
            \nobreak\hskip\z@ plus\fontdimen3\font minus\fontdimen4\font
            \discretionary{\copy\Wrappedvisiblespacebox}{\Wrappedafterbreak}
            {\kern\fontdimen2\font}%
        }%
        \Wrappedbreaksatspecials
        \OriginalVerbatim[#1,codes*=\Wrappedbreaksatpunct]%
    }
    \makeatother

\begin{document}

%
\title{Unitary Encoding of Thermal States via Thermofield Dynamics on Quantum Computers}
\author{G. X. A. Petronilo}
\affiliation{
Universidade Federal do Pará, Salinópolis,PA, Brazil }
\email{gustavopetronilo@gmail.com}

\author{M. R. Araújo}
\affiliation{Latin American Quantum Computing Center, SENAI CIMATEC, Salvador, Brazil}

\author{A. B. M. Souza}
\email{anna.macedo@fbter.org.br}
\affiliation{Latin America Quantum Computing Center, SENAI CIMATEC, Salvador, Brasil.}

\author{Clebson Cruz}
\email{clebson.cruz@ufob.edu.br}
\affiliation{Centro das Ciências Exatas e das Tecnologias, Universidade Federal do Oeste da Bahia, Rua Bertioga 892, Morada Nobre, 47810-059}

\date{\today}

\begin{abstract}
Quantum computing has attracted the attention of the scientific community in the past few decades. However, despite some relevant advantages, near-term quantum devices remain severely limited by thermal effects, which induce decoherence and restrict coherent control at finite temperature.
In this regard, this work reports a gate-based quantum algorithm that prepares the finite-temperature vacuum of Thermofield Dynamics (TFD) and tracks its real-time evolution.
The circuit depth scales linearly with system size and requires only single-qubit rotations and nearest-neighbor CNOT gates, making it NISQ-friendly.
We benchmark the protocol on the PennyLane simulator: magnetization of a spin-$1/2$ particle in a magnetic field agrees with the exact result $M(\beta)=\tanh(\beta\omega/2)$ to machine precision, and the coherent precession acquires a temperature-dependent damping that quantitatively matches the analytical TFD prediction.
Our work provides a ready-to-deploy toolbox for thermal quantum simulations and opens a route to study dissipative phase transitions, quantum thermodynamics and thermal machine-learning models on near-term devices.
\end{abstract}
\label{page:firstblob}

\maketitle

\section{Introduction}

Thermal effects constitute a substantial challenge in quantum computing, both fundamentally and in practice. Finite temperature effects and the inevitable interaction of qubits with their surrounding environment give rise to decoherence processes, thereby impairing the fidelity and robustness of quantum information processing \cite{schlosshauer2019quantum}. These limitations are particularly severe on noisy intermediate-scale quantum (NISQ) devices, where qubit number and coherence times are severely constrained \cite{doi:10.1021/acsomega.3c09720,gyongyosi2020decoherence,ippoliti2020manybody}. Therefore, there is strong motivation to develop strategies for incorporating thermal effects into quantum simulations—approaches that avoid ensemble sampling and instead encode thermal states into single pure quantum states~\cite{PhysRevA.102.022622,he2025tpq,iwaki2022purity,liu2024deep}. Such methods promise to exploit quantum hardware to access finite-temperature phenomena with reduced overhead, offering a practical route for near-term studies of quantum thermodynamics.

On the other hand, the doubled-space formalism of TFD has appeared in the literature as a practical framework for addressing finite-temperature problems in quantum information theory \cite{Takahashi1975,Umezawa1982,Ojima1981,Santana1999,Khanna2009}. 
Rather than working with a mixed Gibbs state, TFD introduces a ``tilde'' copy of the system and applies a temperature-dependent Bogoliubov rotation to build a pure entangled state whose physical reduced density matrix is exactly the canonical ensemble \cite{Umezawa1982}.
Because all thermal averages reduce to expectation values in this doubled vacuum, the formalism sidesteps the need for statistical sampling; moreover, the time evolution at finite temperature is encoded as unitary dynamics generated by the difference Hamiltonian $H-\widetilde H$ \cite{Khanna2009}.

Consequently, TFD has been widely applied across several areas: in gauge field theory, , it provided a consistent operator-based treatment of thermal gauge and matter fields \cite{Ojima1981,Santana1999}; in condensed-matter , it has been employed to study thermal correlation functions and phase transitions within an operator formalism \cite{Khanna2009}; Applications in quantum thermodynamics and open quantum systems have also been explored, where dissipative processes are modeled through enlarged unitary dynamics \cite{hardman1990thermodynamics}; simulations on quantum computers with gate-based and variational algorithms to prepare TFD sttes on quantum hardware \cite{petronilo2021simulating, zhu2020generation, miceli2019thermo, qian2024quantum}, and experimental demonstrations on trapped-ion and superconducting platforms have highlighted the feasibility of doubled-space thermal simulations on real devices \cite{zhu2020generation,Wu2019}.
The doubled-space paradigm has been extended to quantum machine learning, notably in the construction of quantum Boltzmann machines based on thermofield representations of Gibbs states \cite{LulaRocha2023}. In parallel, thermofield  doubled-space have emerged as a key resource in digital quantum simulations of strongly correlated and gravitational systems, including recent implementations on quantum processors inspired by holographic duality \cite{Jafferis2022}.

In this scenario, this work explores the TFD application on quantum computing and quantum simulations in order to built a a gate-based quantum algorithm that prepares the finite-temperature vacuum of TFD and tracks its real-time evolution. We present a \emph{platform-agnostic, gate-based recipe} for preparing the TFD vacuum on any NISQ device.
Starting from the fermionic Bogoliubov operator $G(\beta)=\theta(a^\dagger\widetilde a^\dagger-a\widetilde a)$, we apply the Jordan–Wigner mapping to obtain a spin generator that contains only nearest-neighbor Pauli strings.
The resulting circuit contains $\mathcal O(n)$ single-qubit rotations and $\mathcal O(n)$ CNOT gates for $n$ modes, and can be executed on today's cloud quantum computers.
We implement the protocol in PennyLane and benchmark it on the exactly solvable case of a single spin-1/2 particle in a magnetic field: the magnetization matches the analytical result $M(\beta)=\tanh(\beta\omega/2)$ to machine precision, and coherent precession exhibits the temperature-dependent damping predicted by TFD. Our results provides a ready-to-deploy framework for thermal quantum simulations, thereby establishing a systematic approach to investigate dissipative phase transitions, quantum thermodynamic phenomena, and thermally driven quantum machine-learning models on near-term quantum hardware.

\section{ThermoField Dynamics for Spin Systems}

ThermoField Dynamics (TFD) provides a Hamiltonian-based formalism in which thermal expectation values are represented as expectation values on a \emph{pure} state in an enlarged Hilbert space. 
Given a quantum operator $a$ on the system Hilbert space $\mathcal{S}$, one defines a corresponding tilde operator $\widetilde{a}$ acting on a fictitious Hilbert space copy $\widetilde{\mathcal{S}}$ (the thermal or tilde space) \cite{petronilo2021simulating}.
For each physical fermionic mode, an auxiliary “tilde’’ mode is introduced, forming a doubled Hilbert space 
\(\mathcal{H} \otimes \widetilde{\mathcal{H}}\).  
Thermal properties are encoded through correlations between the real and tilde sectors, and temperature enters via a Bogoliubov rotation parametrized by the inverse temperature \(\beta = 1/T\).

\subsection{Fermionic System and Thermal Bogoliubov Transformation}

For a single fermionic mode of energy $\omega$, the TFD construction starts
from the doubled vacuum state $|0\widetilde{0}\rangle$, which is annihilated by both the physical
annihilation operator $a$ and its tilde counterpart $\widetilde a$
\cite{Takahashi1975,Umezawa1982,Khanna2009}.
The introduction of the auxiliary tilde mode enlarges the Hilbert space such that thermal mixed
states can be represented as pure states in the doubled space.

Finite temperature is incorporated through a \emph{thermal Bogoliubov transformation} \cite{PhysRevD.97.045015,Vdovin2010,ELMFORS1994577} that
coherently mixes the vacuum sector $|0\widetilde{0}\rangle$ with the doubly excited sector
$|1\widetilde{1}\rangle$, generating an entangled state whose reduced density matrix reproduces the
canonical Gibbs ensemble of the physical fermionic mode
\cite{Umezawa1982,Ojima1981,Santana1999}.
The resulting thermal vacuum state can be written as
\begin{equation}
|0(\beta)\rangle
= \cos\theta \, |0\widetilde{0}\rangle
+ \sin\theta \, |1\widetilde{1}\rangle
= U(\beta)\,|0\widetilde{0}\rangle,
\end{equation}
where the thermal angle $\theta$ satisfies
\begin{equation}
\tan\theta = e^{-\beta\omega/2},
\end{equation}
being fixed by the Boltzmann weight associated with the excitation
energy. This choice ensures that the probability of occupying the excited state is
$e^{-\beta\omega}/(1+e^{-\beta\omega})$, in agreement with the Fermi--Dirac distribution for a
single fermionic level \cite{Khanna2009}. From a physical perspective, the Bogoliubov transformation in TFD plays the role of a thermalization mechanism: it generates entanglement between the original system and an auxiliary (tilde) copy in such a way that the reduced state of the original system is a thermal (Gibbs) state \cite{petronilo2021simulating}.

In this regard, the Bogoliubov transformation is implemented by the unitary operator $U(\beta)=e^{G(\beta)}$, generated by the anti-Hermitian operator
\begin{equation} \label{eq:gen_G}
G(\beta) 
= \theta(\beta)\left(a^\dagger \widetilde{a}^\dagger - a\, \widetilde{a}\right),
\qquad
G^\dagger(\beta) = -G(\beta).
\end{equation}
The structure of \(G(\beta)\) encodes the creation and annihilation of pairs across the real--tilde boundary, which in turn produces exactly the entanglement required to represent thermal statistics. This pairing mechanism generates precisely the minimal entanglement required for purification:
tracing out the tilde degree of freedom yields the thermal density operator of the physical mode,
while the full dynamics remains unitary in the doubled Hilbert space
\cite{Umezawa1982,Khanna2009}. 

In practical terms, implementing $U(\beta)$ on a quantum computer would mean coupling a system qubit with an ancillary qubit (representing the tilde system) in a specific way determined by $\beta$. In particular, it has been demonstrated that $U(\beta)$ can be realized via a compact two-qubit gate sequence consisting of a single-axis rotation followed by an entangling gate \cite{petronilo2021simulating}.

In particular, the Bogoliubov transformation method avoids the need for variational optimization of a thermal ansatz (unlike some alternative approaches), resulting in significantly fewer quantum gates to prepare $|0(\beta)\rangle$. This reduction in circuit depth is crucial for noisy intermediate-scale quantum (NISQ) devices and has been demonstrated to make finite-temperature quantum simulations more feasible on current hardware \cite{zhu2020generation, miceli2019thermo}.

\subsection{Jordan--Wigner Transformation for Spin Systems}

In order to implement the TFD construction on a qubit-based quantum computer, fermionic operators first be mapped to spin operators acting on qubits. The standard method for achieving this mapping is the Jordan--Wigner (JW) transformation, which provides an exact correspondence between fermionic creation/annihilation operators and tensor products of Pauli
matrices.

For a system with a single fermionic mode, the JW transformation simplifies considerably. The fermionic annihilation and creation operators, $a$ and $a^\dagger$, can be represented in terms of spin ladder operators, which are themselves expressed through Pauli matrices:
\begin{equation} \label{eq:Jordan_wigner}
a = S^+ = \frac{X + iY}{2},
\qquad
a^\dagger = S^- = \frac{X - iY}{2}.
\end{equation}
Here $X$, $Y$, and $Z$ denote the usual Pauli matrices acting on the physical qubit. The operators $S^+$ and $S^-$ raise and lower the qubit occupation, respectively, reproducing the fermionic action
\(
a|0\rangle = 0,\;
a^\dagger|0\rangle = |1\rangle.
\)

In the TFD formalism, each physical degree of freedom is accompanied by an auxiliary copy (the tilde sector), representing the thermal bath. This doubling of the Hilbert space introduces a second qubit with operators
$\widetilde{X},\widetilde{Y},\widetilde{Z}$ and corresponding ladder operators
\(
\widetilde{S}^\pm = (\widetilde{X} \pm i\widetilde{Y})/2.
\)

A crucial feature of fermionic systems is the anticommutation relations
\begin{equation}
\{a,a^\dagger\}=1,
\qquad
\{a,\widetilde{a}\}=0,
\qquad
\{a,\widetilde{a}^\dagger\}=0.
\end{equation}
To ensure that operators acting on different fermionic modes anticommute correctly when mapped to spins, the Jordan--Wigner transformation introduces  nonlocal phase factor known as the \emph{JW string}. In a system with two modes (the physical and tilde sectors), the operators for the tilde mode acquire an additional $Z$ operator acting on the first qubit:
\begin{equation}
\widetilde{a} = Z\,\widetilde{S}^+,
\qquad
\widetilde{a}^\dagger = Z\,\widetilde{S}^-.
\end{equation}

Although the present system contains only a single physical mode and its tilde copy, the JW string remains essential. Without this $Z$ factor, operators from the two sectors would commute rather than anticommute, violating the fermionic algebra. The JW string therefore preserves the correct fermionic statistics in the qubit representation.

\subsection{Simplification Using Spin Relations}

After applying the Jordan–Wigner transformation, the generator of TFD transformation can be expressed in terms of spin ladder operators acting on the physical and auxiliary qubits. At this stage it is convenient to exploit simple algebraic identities involving Pauli matrices in order to simplify the resulting expression.
Using the Pauli matrix relations
\[ZX=-XZ,\qquad ZY=-YZ,\]
one obtains the identities
\[
S^+ Z = -S^+,
\qquad
S^- Z = S^-.
\]
These relations allow the Jordan–Wigner string appearing in the tilde-sector operators to be absorbed into the ladder operators when constructing the TFD generator. As a result, the generator simplifies considerably and can be written purely in terms of spin ladder operators acting on the two qubits, 
\begin{equation}
G
= -\theta \left(S^- \widetilde{S}^- + S^+ \widetilde{S}^+\right).
\end{equation}

Expressing the ladder operators in terms of Pauli matrices yields the compact Pauli form:
\begin{equation}
G
= -\frac{\theta}{2}\left(X\widetilde{X} - Y\widetilde{Y}\right).
\end{equation}

This expression has an important practical advantage: it involves only two-qubit Pauli interactions, which are directly implementable on quantum hardware using standard gate decompositions.

However, one must also ensure that the generator produces a unitary transformation when exponentiated. Since this expression is Hermitian, while the original fermionic generator was anti-Hermitian, we introduce a factor of \(i\) to preserve the correct structure of the transformation. The appropriate generator for the spin representation is therefore
The correct spin-system generator is therefore
\begin{equation}
G_{\text{spin}}
= -i\frac{\theta}{2}\left(X\widetilde{X} - Y\widetilde{Y}\right),
\qquad
U = e^{G_{\text{spin}}}.
\end{equation}

Acting with this unitary on \(|0\widetilde{0}\rangle\) gives the thermal vacuum in the spin representation:
\begin{equation}
e^{G_{\text{spin}}}\, |0\widetilde{0}\rangle
= \cos\theta\,|0\widetilde{0}\rangle
- i \sin\theta\,|1\widetilde{1}\rangle.
\end{equation}
This two-qubit entangled state is the TFD thermal vacuum encoded directly in computational-basis qubits, suitable for quantum simulation.

\section{Theoretical Analysis of Observables}

\subsection{Magnetization in the TFD Formalism}
In TFD framework, the magnetization $M$ is introduced as the thermal expectation value of the total spin operator along the $z$-direction, evaluated in the thermal vacuum state,
\begin{equation}
M = \langle 0(\beta) | \sum_i Z_i | 0(\beta) \rangle.
\end{equation}
For a single spin, this expectation value can be written as
\begin{equation}
M = \cos 2\theta= \tanh\Big(\frac{\omega\beta}{2}\Big),
\end{equation}
This result shows that the magnetization decreases monotonically with temperature, approaching unity at low temperature and vanishing in the high-temperature limit.

\subsection{Spin Precession with Thermal Effects}
We now analyze the dynamics of a spin-$1/2$ system in a magnetic field, focusing on how thermal effects modify its precessional behavior.

Consider a spin-$1/2$ particle placed in a static magnetic field $B$ along the $z$-axis. The system is described by the Hamiltonian:
\begin{equation}
H = -\frac{\omega}{2} Z, \quad \omega = \mu B.
\end{equation}
The ground state $|0\rangle$ has energy $-\omega/2$, while $|1\rangle$ has $\omega/2$.

Under unitary time evolution generated by $H$, the system undergoes Larmor precession with frequency $\omega$, corresponding to coherent oscillations between the spin components.

Whe then proceed to incorporate thermal effects within the real-time framework of TFD. The time evolution operator is generated by  the Hamiltonian proportional to the difference of the physical and tilde operators,
\begin{equation*}
    H_T = H - \widetilde{H}.
\end{equation*}
This structure ensures that the dynamics of the auxiliary system mirrors that of the physical system, but with an opposite sign, which is a key feature of the formalism.

For the system under consideration, where the Hamiltonian is proportional to the operator $Z$, the total generator becomes proportional to $Z- \widetilde{Z}$. The corresponding time evolution operator is therefore given by
\begin{equation}
T(t) = \exp\left(i\omega t (Z - \widetilde{Z})\right),
\end{equation}
which produces opposite phase rotations in the original and auxiliary degrees of freedom.

Now let ${P}(\beta) = U^\dagger(\beta) P U(\beta)$ denote a Bogoliubov-transformed operator. Its time evolution is:
\begin{equation}
P(t, \beta) = T^\dagger(t) P(\beta) T(t).
\end{equation}
For Pauli operators, we derive:
\begin{enumerate}
    \item $X(t, \beta)$:
    \begin{align*}
    {X}(t, \beta) = \cos\theta \left(\cos(\omega t) X - \sin(\omega t) Y\right) - \\ \sin\theta \left(Z \left(\cos(\omega t) \widetilde{Y} + \sin(\omega t) \widetilde{X}\right)\right)
    \end{align*}
    
    \item ${Y}(t, \beta)$:
    \begin{align*}
    {Y}(t, \beta) = \cos\theta \left(\cos(\omega t) Y + \sin(\omega t) X\right) - \\ 
    -\sin\theta \left(Z \left(\cos(\omega t) \widetilde{X} - \sin(\omega t) \widetilde{Y}\right)\right)
    \end{align*}
    
    \item ${Z}(t, \beta)$:
    \begin{align*}
    {Z}(t, \beta) &= \frac{\cos 2\theta}{2} (Z + \widetilde{Z}) + \frac{\sin 2\theta}{2}  \left[(\cos(\omega t) Y \right. \\ &\left. + \sin(\omega t) X)(\cos(\omega t) \widetilde{X} - \sin(\omega t) \widetilde{Y})\right] + \frac{1}{2} (Z - \widetilde{Z})
    \end{align*}
\end{enumerate}

In this regard, we can evaluate the expectation values of the Pauli operators for different initial states within the TFD framework. 
For the states $|0\widetilde{0}\rangle$ and $|1\tilde{1}\rangle$:
\begin{equation*}
\langle X(t,\beta) \rangle = \langle Y(t,\beta) \rangle = 0,
\end{equation*}
while the longitudinal magnetization is given by
\begin{equation*}
\langle Z(t,\beta) \rangle = \pm \cos(2\theta),
\end{equation*}
where the sign depends on the chosen state. This reflects the fact that these states are eigenstates of the $Z$ operator and, therefore, do not exhibit coherent oscillations under the considered dynamics.

In contrast, for the superposition state $|+\widetilde{0}\rangle$, nontrivial coherent dynamics emerges. The expectation values of the transverse components exhibit oscillatory behavior,
\begin{align*}
\langle X(t,\beta) \rangle &= \cos\theta \cos(\omega t), \\
\langle Y(t,\beta) \rangle &= \cos\theta \sin(\omega t),
\end{align*}
while the longitudinal component becomes
\begin{equation*}
\langle Z(t,\beta) \rangle = \frac{1}{2}\Big(\cos(2\theta)-1\Big).
\end{equation*}
The cosine factor directly controls the amplitude of the oscillations, indicating that thermal effects can suppress quantum coherence as temperature increases.

    
Using these results in terms of the reciprocal temperature $\beta$, using the relation $\tan\theta = e^{-\beta\omega/2}$, the expectation values assume a more evident thermodynamic interpretation. For the states $|0\widetilde{0}\rangle$ and $|1\widetilde{1}\rangle$, one obtains
\begin{equation*}
\langle Z(t,\beta) \rangle = \pm \tanh\!\left(\frac{\omega\beta}{2}\right), 
\qquad
\langle X(t,\beta) \rangle = \langle Y(t,\beta) \rangle = 0,
\end{equation*}
which corresponds to the thermal equilibrium magnetization of a two-level system.

On the other hand, for the superposition state, the transverse components become
\begin{align*}
\langle X(t,\beta) \rangle &= \frac{e^{\omega\beta/2}}{\sqrt{e^{\omega\beta}+1}}\, \cos(\omega t), \\
\langle Y(t,\beta) \rangle &= \frac{e^{\omega\beta/2}}{\sqrt{e^{\omega\beta}+1}}\, \sin(\omega t),
\end{align*}
while the longitudinal component reads
\begin{equation*}
\langle Z(t,\beta) \rangle = \frac{1}{2}\left[\tanh\!\left(\frac{\omega\beta}{2}\right)-1\right].
\end{equation*}

These expressions make explicit how temperature modulates both equilibrium properties and dynamical behavior: as $\beta$ decreases (higher temperatures), the transverse amplitudes are suppressed and the system progressively loses coherence, consistently with the expected thermal damping encoded through the TFD formalism.

%
%

\section{Quantum Algorithm Implementation}
The numerical implementation of TFD formalism in this paper was done using the Xanadu PennyLane framework \cite{bergholm2018pennylane}, which provides native tools for quantum circuit simulation and fermionic operator manipulation. For reasons of reproducibility, we present in this section the step-by-step algorithm and its implementation.

\subsection{Auxiliary Functions}
The first step consists of importing  all the necessary libraries, including PennyLane for quantum simulation \cite{bergholm2018pennylane}, NumPy for numerical operations \cite{harris2020array}, and SciPy for matrix exponentiation \cite{2020SciPy-NMeth}, as shown in Box 1.1. In addition, we make use of PennyLane's fermionic module \cite{arrazola2023}, which simplifies the manipulation of creation and annihilation operators.

\begin{tcolorbox}
    [breakable, size=fbox, boxrule=1pt, pad at break*=1mm, colback=cellbackground, colframe=framec, coltitle=black, title=Box 1.1: Importing the libraries]
    \begin{Verbatim}[commandchars=\\\{\}]
\textcolor{PY_green}{import} pennylane \textcolor{PY_green}{as} qml
\textcolor{PY_green}{from} pennylane \textcolor{PY_green}{import} numpy \textcolor{PY_green}{as} np
\textcolor{PY_green}{from} pennylane.fermi \textcolor{PY_green}{import} FermiA, FermiC
\textcolor{PY_green}{import} matplotlib.pyplot \textcolor{PY_green}{as} plt
\textcolor{PY_green}{from} scipy.linalg \textcolor{PY_green}{import} expm
    \end{Verbatim}
\end{tcolorbox}

For both routines of studying the TFD magnetization and its time evolution, it is necessary to prepare the tilde-conjugate operators and the generator $G(\beta)$ as made explicit in equation \ref{eq:gen_G}. This process can be computationally defined as the function in Box 1.2. PennyLane built-in classes \texttt{FermiA} and \texttt{FermiC} facilitate the implementation of annihilation and creation of fermionic operators. The Jordan-Wigner mapping described by equation \ref{eq:Jordan_wigner} is then applied through the use of \texttt{jordan\_wigner}, converting the fermionic operators into qubit operators.

\begin{tcolorbox}
    [breakable, size=fbox, boxrule=1pt, pad at break*=1mm, colback=cellbackground, colframe=framec, coltitle=black, title=Box 1.2: Defining the precompute operators function.]
    \begin{Verbatim}[commandchars=\\\{\}]
\textcolor{PY_green}{def} \textcolor{PY_blue}{precompute_operators}(n_modes):
    G_ops = []
    \textcolor{PY_green}{for} i \textcolor{PY_green}{in range}(n_modes):
        a_i = FermiA(i)
        a_i_dag = FermiC(i)
        tilde_a_i = FermiA(i + n_modes)
        tilde_a_i_dag = FermiC(i + n_modes)
        G = -\textcolor{PY_green}{1} * (a_i * tilde_a_i - a_i_dag * tilde_a_i_dag)
        G_ops.\textcolor{PY_blue}{append}(qml.\textcolor{PY_blue}{jordan_wigner}(G, ps = \textcolor{PY_green}{False}))
    \textcolor{PY_green}{return} G_ops
    \end{Verbatim}
\end{tcolorbox}

This function plays a central role in the algorithm, as it provides the operational interface between the abstract TFD formalism and its concrete implementation on a qubit-based quantum simulator.

\subsection{Magnetization Simulation}

With the operators defined, the next step is to construct a quantum circuit that prepares the thermal vacuum state and evaluates the physical observables. In particular, we focus on the magnetization components, which serve as thermodynamic indicators of the system. The function below, presented in Box 1.3, implements the preparation of the TFD state by applying an unitary transformation, where the thermal angle $\theta$ is calculated according to the analytical expression derived in the previous section. Thus, the \texttt{tfd\_circuit} function utilizes the definition of $\theta$ as in Eq. \ref{eq:gen_G} and $G$ to establish the Hamiltonian.


\begin{tcolorbox}
    [breakable, size=fbox, boxrule=1pt, pad at break*=1mm, colback=cellbackground, colframe=framec, coltitle=black, title=Box 1.3: Getting the thermal state.]
    \begin{Verbatim}[commandchars=\\\{\}]
\textcolor{PY_green}{def} \textcolor{PY_blue}{get_thermal_state}(beta, omega, n_modes=1):
    n_wires = 2 * n_modes

    G_ops = precompute_operator(n_modes)

    dev = qml.\textcolor{PY_blue}{device}(\textcolor{PY_red}{"default.qubit"}, wires = n_wires)

    @qml.qnode(dev)
    \textcolor{PY_green}{def} \textcolor{PY_blue}{tfd_circuit}(beta):
        \textcolor{PY_green}{for} i \textcolor{PY_green}{in range}(n_modes):
            theta = np.\textcolor{PY_blue}{arctan}(np.\textcolor{PY_blue}{exp}(- beta * omega / \textcolor{PY_green}{2}))
            H = qml.\textcolor{PY_blue}{Hamiltonian}([\textcolor{PY_green}{1.0}], [G_ops[i]])
            qml.\textcolor{PY_blue}{ApproxTimeEvolution}(H, theta, \textcolor{PY_green}{1})

        m_x = \textcolor{PY_green}{sum}(qml.\textcolor{PY_blue}{PauliX}(i) \textcolor{PY_green}{for} i \textcolor{PY_green}{in range}(n_modes))
        m_y = \textcolor{PY_green}{sum}(qml.\textcolor{PY_blue}{PauliY}(i) \textcolor{PY_green}{for} i \textcolor{PY_green}{in range}(n_modes))
        m_z = \textcolor{PY_green}{sum}(qml.\textcolor{PY_blue}{PauliZ}(i) \textcolor{PY_green}{for} i \textcolor{PY_green}{in range}(n_modes))

        \textcolor{PY_green}{return} qml.\textcolor{PY_blue}{expval}(m_x), qml.\textcolor{PY_blue}{expval}(m_y), qml.\textcolor{PY_blue}{expval}(m_z)

    \textcolor{PY_green}{return} tfd_circuit(beta)
    \end{Verbatim}
\end{tcolorbox}

Therefore, in this implementation, the operator $G(\beta)$ is promoted to a Hamiltonian and its exponential is approximated using a first-order time evolution operator. This effectively realizes the Bogoliubov transformation as a quantum gate sequence.

After preparing the state, the expectation values of the Pauli operators are computed, yielding the magnetization components $(m_x, m_y, m_z)$.

The physical parameters of the system are defined in the following Box 1.4. In particular, we scan a range of inverse temperatures $\beta$ to analyze the thermal behavior of the system.


\begin{tcolorbox}
    [breakable, size=fbox, boxrule=1pt, pad at break*=1mm, colback=cellbackground, colframe=framec, coltitle=black, title=Box 1.4: Setting the system parameters.]
    \begin{Verbatim}[commandchars=\\\{\}]
omega = 0.5
n_modes = 1
betas = np.\textcolor{PY_blue}{linspace}(0.1, 5.0, 50)
    \end{Verbatim}
\end{tcolorbox}

Then, the magnetization is then calculated from Box 1.5 for each value of $\beta$, allowing for a direct comparison between numerical results and the theoretical predictions.

\begin{tcolorbox}
    [breakable, size=fbox, boxrule=1pt, pad at break*=1mm, colback=cellbackground, colframe=framec, coltitle=black, title=Box 1.5: Getting the magnetization.]
    \begin{Verbatim}[commandchars=\\\{\}]
m_x_vals, m_y_vals, m_z_vals = \textcolor{PY_green}{zip}(*[prepare_tfd_state(beta, omega, n_modes) \textcolor{PY_green}{for} beta \textcolor{PY_green}{in} betas])

theoretical_z = np.\textcolor{PY_blue}{tanh}(betas * omega / \textcolor{PY_green}{2})
    \end{Verbatim}
\end{tcolorbox}

Finally, the results can be visualized using Box 1.6, showing the dependence of the magnetization components on the inverse temperature. As a comparative measure, the agreement with the analytical curve $\tanh(\beta \omega /2)$ is used to validate the correctness of the proposed implementation.

\begin{tcolorbox}
    [breakable, size=fbox, boxrule=1pt, pad at break*=1mm, colback=cellbackground, colframe=framec, coltitle=black, title=Box 1.6: Plotting the results]
    \begin{Verbatim}[commandchars=\\\{\}]
plt.\textcolor{PY_blue}{figure}(figsize=(\textcolor{PY_green}{10}, \textcolor{PY_green}{6}))
plt.\textcolor{PY_blue}{plot}(betas, m_x_vals, \textcolor{PY_red}{'r-'}, label=\textcolor{PY_red}{r'$\textbackslash langle m_x \textbackslash rangle$'})
plt.\textcolor{PY_blue}{plot}(betas, m_y_vals, \textcolor{PY_red}{'g-'}, label=\textcolor{PY_red}{r'$\textbackslash langle m_y \textbackslash rangle$'})
plt.\textcolor{PY_blue}{plot}(betas, m_z_vals, \textcolor{PY_red}{'b-'}, label=\textcolor{PY_red}{r'$\textbackslash langle m_z \textbackslash rangle$'})
plt.\textcolor{PY_blue}{plot}(betas, theoretical_z, \textcolor{PY_red}{'k--'}, label= \textcolor{PY_red}{r'Theoretical $\textbackslash langle m_z \textbackslash rangle$'})
plt.\textcolor{PY_blue}{xlabel}(\textcolor{PY_red}{r'Inverse Temperature $ \textbackslash beta$'})
plt.\textcolor{PY_blue}{ylabel}(\textcolor{PY_red}{'Magnetization'})
plt.\textcolor{PY_blue}{title}(\textcolor{PY_red}{f'Magnetization Components vs. Inverse Temperature ($\textbackslash \textbackslash omega={omega}$, $n={n_modes}$)'})
plt.\textcolor{PY_blue}{legend}()
plt.\textcolor{PY_blue}{grid}()
plt.\textcolor{PY_blue}{show}()
    \end{Verbatim}
\end{tcolorbox}

\subsection{Time Evolution Simulation}

As the next step of the algorithm, we extend the analysis to the dynamical regime by studying the time evolution of the TFD state. In this regard, Box 2.1 shows the procedure for preparing the thermal state to be used in the subsequent time evolution. The initial step remains the preparation of the thermal vacuum, but now the full quantum state is explicitly returned for subsequent evolution.

\begin{tcolorbox}
    [breakable, size=fbox, boxrule=1pt, pad at break*=1mm, colback=cellbackground, colframe=framec, coltitle=black, title=Box 2.1: Getting the thermal state for time evolution.]
    \begin{Verbatim}[commandchars=\\\{\}]
\textcolor{PY_green}{def} \textcolor{PY_blue}{get_thermal_state}(beta, omega, n_modes=1):
    n_wires = 2 * n_modes

    G_ops = precompute_operator(n_modes)

    dev = qml.\textcolor{PY_blue}{device}(\textcolor{PY_red}{"default.qubit"}, wires = n_wires)

    @qml.qnode(dev)
    \textcolor{PY_green}{def} \textcolor{PY_blue}{thermal_state_circuit}(beta):
        \textcolor{PY_green}{for} i \textcolor{PY_green}{in range}(n_modes):
            qml.\textcolor{PY_blue}{Hadamard}(wires=i)
        \textcolor{PY_green}{for} i \textcolor{PY_green}{in range}(n_modes):
            theta = np.\textcolor{PY_blue}{arctan}(np.\textcolor{PY_blue}{exp}(- beta * omega / \textcolor{PY_green}{2}))
            H = qml.\textcolor{PY_blue}{Hamiltonian}([theta], [G_ops[i]])
            qml.\textcolor{PY_blue}{QubitUnitary}(expm(-1j * H.\textcolor{PY_blue}{matrix}()), wires = \textcolor{PY_green}{range}(n_wires))
        \textcolor{PY_green}{return} qml.\textcolor{PY_blue}{state}()

    \textcolor{PY_green}{return} thermal_state_circuit(beta)
    \end{Verbatim}
\end{tcolorbox}

In contrast to the previous subroutine, the dynamical implementation constructs the exact unitary operator using matrix exponentiations. This  procedure allows a direct representation of the Bogoliubov transformation. Once the initial state is prepared, we simulate its time evolution under a given Hamiltonian. The evolution is implemented by applying the unitary operator $e^{-iHt}$ at discrete time steps. For the observation of the dynamics of the system, some characteristics are defined in the same way, such as the Hamiltonian. In this context, the function defined in Box 2.2 performs this task by combining three essential steps: (i) state initialization, (ii) unitary time evolution, and (iii) measurement of observables.  Therefore, this function initializes the quantum state, applies the time evolution operator, and computes the expectation values of the magnetization observables at each time step.

\begin{tcolorbox}
    [breakable, size=fbox, boxrule=1pt, pad at break*=1mm, colback=cellbackground, colframe=framec, coltitle=black, title=Box 2.2: Evolving the TFD.]
    \begin{Verbatim}[commandchars=\\\{\}]
\textcolor{PY_green}{def} \textcolor{PY_blue}{evolve_tfd_state}(state, H, time_steps, n_modes=1):
    n_wires = \textcolor{PY_green}{int}(np.\textcolor{PY_blue}{log2}(\textcolor{PY_green}{len}(state)))
    dev = qml.\textcolor{PY_blue}{device}(\textcolor{PY_red}{"default.qubit"}, wires=n_wires)
    
    
    observables = [
        \textcolor{PY_green}{sum}(qml.\textcolor{PY_blue}{PauliX}(i) \textcolor{PY_green}{for} i \textcolor{PY_green}{in range}(n_modes)),  
        \textcolor{PY_green}{sum}(qml.\textcolor{PY_blue}{PauliY}(i) \textcolor{PY_green}{for} i \textcolor{PY_green}{in range}(n_modes)), 
        \textcolor{PY_green}{sum}(qml.\textcolor{PY_blue}{PauliZ}(i) \textcolor{PY_green}{for} i \textcolor{PY_green}{in range}(n_modes))   
    ]
    
    @qml.qnode(dev)
    \textcolor{PY_green}{def} \textcolor{PY_blue}{time_evolution_circuit}(t):
        qml.\textcolor{PY_blue}{StatePrep}(state, wires=range(n_wires))
        qml.\textcolor{PY_blue}{QubitUnitary}(expm(-\textcolor{PY_green}{1j} * H.\textcolor{PY_blue}{matrix}() * t), wires=\textcolor{PY_green}{range}(n_wires))
        \textcolor{PY_green}{return} [qml.expval(o) \textcolor{PY_green}{for} o \textcolor{PY_green}{in} observables]
    
    results = []
    \textcolor{PY_green}{for} t \textcolor{PY_green}{in} time_steps:
        results.\textcolor{PY_blue}{append}(time_evolution_circuit(t))
    
    \textcolor{PY_green}{return} np.\textcolor{PY_blue}{array}(results).\textcolor{PY_blue}{T}
    \end{Verbatim}
\end{tcolorbox}

Box 2.3 specifies the initial conditions of the system, including the temperature, system size, and the normalization of the state vector.

\begin{tcolorbox}
    [breakable, size=fbox, boxrule=1pt, pad at break*=1mm, colback=cellbackground, colframe=framec, coltitle=black, title=Box 2.3: Setting the initial values.]
    \begin{Verbatim}[commandchars=\\\{\}]
omega = \textcolor{PY_green}{0.5}
beta = \textcolor{PY_green}{1.0}
n_modes = \textcolor{PY_green}{1}

# Prepare initial state
initial_state = get_thermal_state(beta, omega, n_modes)
initial_state = initial_state/np.\textcolor{PY_blue}{linalg.norm}(initial_state)
    \end{Verbatim}
\end{tcolorbox}

Next, we define the TFD Hamiltonian in terms of number operators for the physical and tilde modes in Box 2.4. The opposite signs reflect the structure of the doubled Hilbert space.

\begin{tcolorbox}
    [breakable, size=fbox, boxrule=1pt, pad at break*=1mm, colback=cellbackground, colframe=framec, coltitle=black, title=Box 2.4: Defining the TFD Hamiltonian and time steps.]
    \begin{Verbatim}[commandchars=\\\{\}]
a_dag_a = qml.\textcolor{PY_blue}{jordan_wigner}(FermiC(\textcolor{PY_green}{0}) * FermiA(\textcolor{PY_green}{0}), ps=\textcolor{PY_green}{False})
tilde_a_dag_tilde_a = qml.\textcolor{PY_blue}{jordan_wigner}(FermiC(\textcolor{PY_green}{1}) * FermiA(\textcolor{PY_green}{1}), ps=\textcolor{PY_green}{False})
H_tfd = qml.Hamiltonian([\textcolor{PY_green}{1.0}, \textcolor{PY_green}{-1.0}], [a_dag_a, tilde_a_dag_tilde_a])


dt = \textcolor{PY_green}{0.1}
total_time = \textcolor{PY_green}{10.0}
time_steps = np.\textcolor{PY_blue}{arange}(\textcolor{PY_green}{0}, total_time, dt)


m_x_vals, m_y_vals, m_z_vals = evolve_tfd_state(initial_state, H_tfd, time_steps)
    \end{Verbatim}
\end{tcolorbox}

Finally, the time evolution results are visualized using Box 2.5. The first plot shows the evolution of all magnetization components for a fixed temperature, while the second plot illustrates how the dynamics of $m_x(t)$ changes across different thermal regimes.

\begin{tcolorbox}
    [breakable, size=fbox, boxrule=1pt, pad at break*=1mm,
     colback=cellbackground, colframe=framec, coltitle=black,
     title=Box 2.5: Plotting the results.]
\begin{Verbatim}[commandchars=\\\{\}]

# --------------------------------------
# Evolution for a single beta
# --------------------------------------
initial_state = get_thermal_state(\textcolor{PY_green}{beta},
                                  \textcolor{PY_green}{omega},
                                  \textcolor{PY_green}{n_modes})
m_x_vals, m_y_vals, m_z_vals = evolve_tfd_state(initial_state,
                                                H_tfd,
                                                time_steps)

plt.\textcolor{PY_blue}{figure}(figsize=(\textcolor{PY_green}{10},
                                         \textcolor{PY_green}{6}))
plt.\textcolor{PY_blue}{plot}(time_steps, m_x_vals,
                              label=\textcolor{PY_red}{"m_x"})
plt.\textcolor{PY_blue}{plot}(time_steps, m_y_vals,
                              label=\textcolor{PY_red}{"m_y"})
plt.\textcolor{PY_blue}{plot}(time_steps, m_z_vals,
                              label=\textcolor{PY_red}{"m_z"})
plt.\textcolor{PY_blue}{xlabel}(\textcolor{PY_red}{"Time"})
plt.\textcolor{PY_blue}{ylabel}(\textcolor{PY_red}{"Magnetization"})
plt.\textcolor{PY_blue}{title}(
    \textcolor{PY_red}{f"Magnetization Components vs Time (beta = \{beta\})"}
)
plt.\textcolor{PY_blue}{grid}()
plt.\textcolor{PY_blue}{legend}()
plt.\textcolor{PY_blue}{show}()

# --------------------------------------
# m_x(t) for multiple beta values
# --------------------------------------
beta_values = [\textcolor{PY_green}{0.01},
               \textcolor{PY_green}{1.0},
               \textcolor{PY_green}{10.0}]

plt.\textcolor{PY_blue}{figure}(figsize=(\textcolor{PY_green}{10},
                                         \textcolor{PY_green}{6}))
for b in beta_values:
    state = get_thermal_state(b,
                              \textcolor{PY_green}{omega},
                              \textcolor{PY_green}{n_modes})
    m_x, _, _ = evolve_tfd_state(state, H_tfd, time_steps)
    plt.\textcolor{PY_blue}{plot}(time_steps, m_x,
                                  label=\textcolor{PY_red}{f"beta = \{b\}"})


plt.\textcolor{PY_blue}{xlabel}(\textcolor{PY_red}{"Time"})
plt.\textcolor{PY_blue}{ylabel}(\textcolor{PY_red}{"m_x"})
plt.\textcolor{PY_blue}{title}(
    \textcolor{PY_red}{"m_x(t) for widely spaced beta values"}
)
plt.\textcolor{PY_blue}{grid}()
plt.\textcolor{PY_blue}{legend}()
plt.\textcolor{PY_blue}{show}()

\end{Verbatim}
\end{tcolorbox}

The obtained results, discussed on next section, demonstrate how the TFD formalism allows one to consistently describe both equilibrium and dynamical properties of quantum systems at finite temperature within the proposed unified quantum circuit framework.

\section{Results and Discussion}

\subsection{Time Evolution and Thermal Damping}
The simulation of the time evolution (Boxes 2.1 to 2.5) of an initial superposition state prepared in the TFD vacuum reveals a clear signature of \emph{thermally-induced coherent damping} \cite{shen2024finite}. The transverse components of magnetization $\langle M_x(t) \rangle$ and $\langle M_y(t) \rangle$ exhibit oscillatory behavior at the Larmor frequency $\omega$, as expected for a spin-$1/2$ system. 
However, in contrast to zero-temperature unitary dynamics, the oscillation amplitude is renormalized by the factor $\cos\theta$, with
\begin{equation}
\cos\theta = \frac{e^{\beta\omega/2}}{\sqrt{e^{\beta\omega}+1}},
\end{equation}
in agreement with the analytical TFD prediction as derived in subsection 3.3.3.

This behavior highlights a key feature of the TFD formalism: thermal effects are encoded as \emph{entanglement-induced amplitude renormalization}, rather than arising from external decoherence channels. 
Since the simulations are performed in an ideal noiseless environment, the observed damping is entirely intrinsic and originates from the Bogoliubov structure of the thermal vacuum. 
This is consistent with previous theoretical and experimental studies showing that TFD states reproduce effective open-system behavior through unitary dynamics in an enlarged Hilbert space \cite{Zhu2020,lee2022variational,qian2024quantum}.

\subsection{Magnetization as a Function of Temperature}

\begin{figure}[H]
    \centering
    \includegraphics[width=\linewidth]{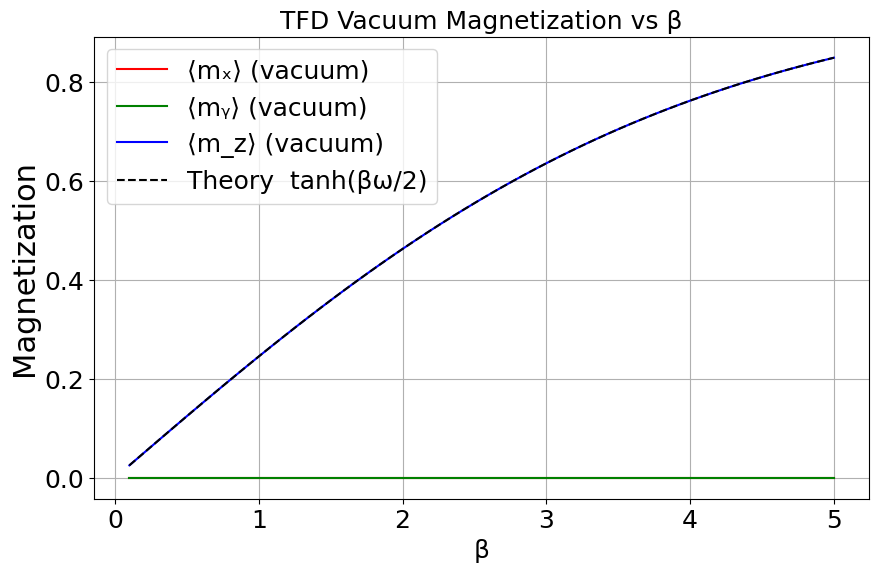}
    \caption{Magnetization $\langle M_z \rangle$ as a function of inverse temperature $\beta$ (blue line) compared to the theoretical result (black dashed line). The $\langle M_x \rangle$ and $\langle M_y \rangle$ components (red and green lines) are zero, as expected for the thermal vacuum state.}
    \label{fig:mz_beta}
\end{figure}

The simulation of the longitudinal magnetization $\langle M_z \rangle$ as a function of the inverse temperature $\beta$ (Boxes 1.3 to 1.6) showed excellent agreement with the exact theoretical result, $M(\beta)=\tanh(\beta\omega/2)$. Figure \ref{fig:mz_beta} illustrates this agreement over the full temperature range.

In the high-temperature limit ($\beta \to 0$), the magnetization approaches zero since the system approaches a maximally mixed state. As $\beta$ increases (lower temperature range), the magnetization approaches the maximum value of $1$, converging to the fully spin-polarized ground state, which is consistent with the zero-temperature limit of the canonical ensemble. The machine precision observed for $\langle M_z \rangle$ validates the correct preparation of the TFD vacuum state $|0(\beta)\rangle$ through the proposed quantum circuit. This level of accuracy is comparable to or exceeds that reported in recent implementations of TFD-based thermal state preparation on quantum simulators and hardware \cite{miceli2019thermo, petronilo2021simulating}. 

This damping is not due to environmental decoherence (which is absent in the ideal simulator) but is an intrinsic effect of the TFD representation of thermal dynamics. The damping factor $\cos\theta$ is a decreasing function of the inverse temperature $\beta$. At $\beta \to \infty$ (zero temperature), $\cos\theta \to 1$, and the precession is fully coherent. At $\beta \to 0$ (infinite temperature), $\cos\theta \to 1/\sqrt{2}$, and the precession amplitude is significantly damped. This result demonstrates the capability of the TFD formalism, when implemented on a quantum computer, to simulate the dynamics of an open system in contact with a thermal reservoir in a unitary manner, without the need for non-unitary super-evolution operators. In particular, while variational approaches typically introduce optimization errors \cite{Wu2019,lee2022variational}, the proposed gate-based construction can achieves exact thermal statistics without the need for classical feedback loops.

\subsection{Magnetization of the Superposition State $|+\widetilde{0}\rangle$}

\begin{figure}[H]
    \centering
    \includegraphics[width=\linewidth]{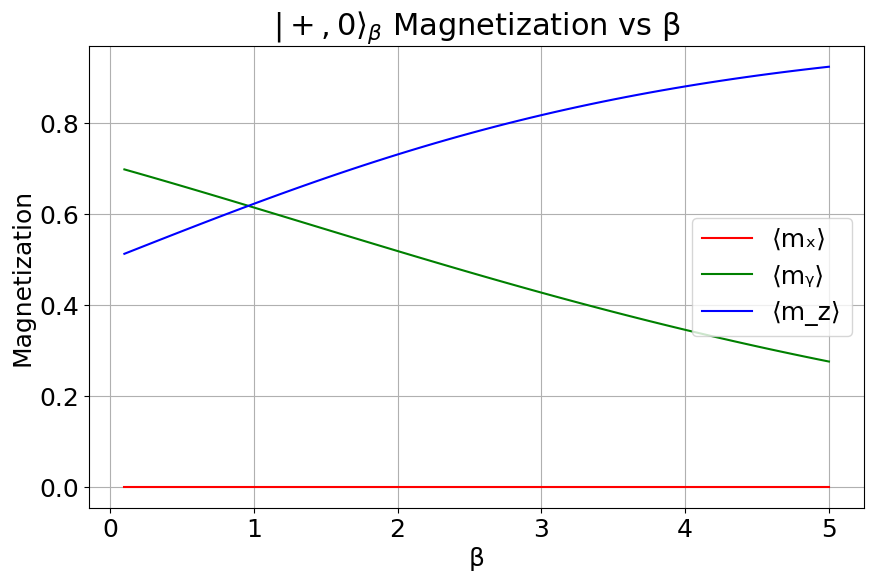}
    \caption{Magnetization components of the superposition state $|+\widetilde{0}\rangle$ at $\beta = 1$. The transverse components $\langle M_x \rangle$ and $\langle M_y \rangle$ show partial amplitude due to thermal damping, while $\langle M_z \rangle$ is negative as expected.}
    \label{fig:plus0_magnetization}
\end{figure}

Figure \ref{fig:plus0_magnetization} shows the magnetization components $\langle M_x \rangle$, $\langle M_y \rangle$, and $\langle M_z \rangle$ for the superposition state $|+\widetilde{0}\rangle$ at inverse temperature $\beta = 1$. The values confirm the thermal reduction in the transverse components and the expected negative value for $\langle M_z \rangle$, in agreement with the theoretical prediction $\langle M_z \rangle = \frac{1}{2} (\tanh(\beta \omega/2) - 1)$, reflecting the thermal suppression of quantum coherence.

These results explicitly demonstrate how thermal fluctuations redistribute population and coherence through the magnetization.  Such behavior is consistent with recent studies on TFD sttes, where finite temperature effectively shrinks the accessible Bloch vector due to entanglement with auxiliary degrees of freedom \cite{Zhu2020, Jafferis2022, Lin2021}.

\subsection{Time Evolution of $|+\widetilde{0}\rangle$ at $\beta = 1$}

\begin{figure}[H]
    \centering
    \includegraphics[width=\linewidth]{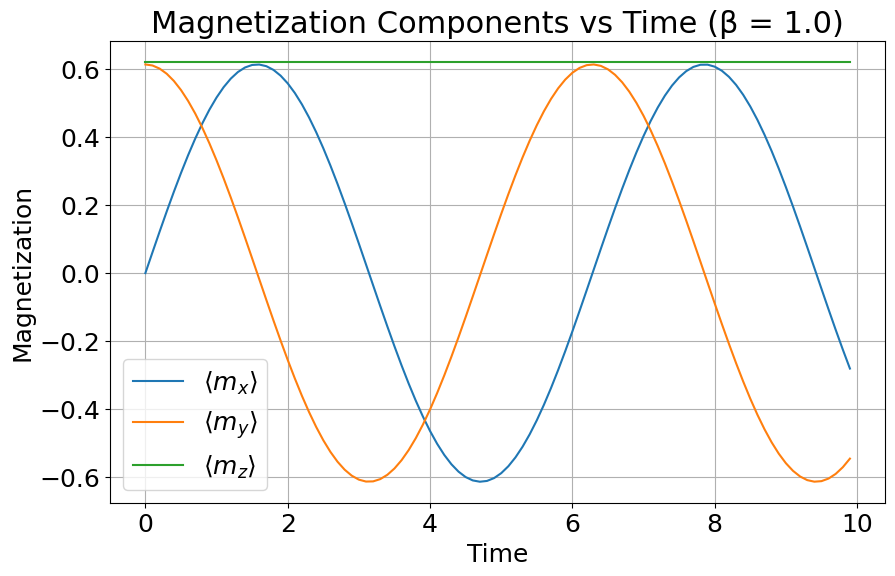}
    \caption{Time evolution of the expectation values $\langle M_x(t) \rangle$, $\langle M_y(t) \rangle$, and $\langle M_z(t) \rangle$ for $|+\widetilde{0}\rangle$ at $\beta = 1$. The transverse oscillations are damped by $\cos\theta$, while the longitudinal component remains negative.}
    \label{fig:plus0_time_evolution}
\end{figure}

The time evolution of the expectation values of the Pauli operators for the state $|+\widetilde{0}\rangle$ at $\beta = 1$ is shown in Figure \ref{fig:plus0_time_evolution}. The transverse components $\langle M_x(t) \rangle$ and $\langle M_y(t) \rangle$ exhibit coherent oscillations at frequency $\omega$, while $\langle M_z(t) \rangle$ remains constant, demonstrating the predicted dynamics of the Bogoliubov-transformed operators.

It is worth noting that, the persistence of coherent oscillations indicates that the system remains fully unitary, despite exhibiting features analogous to dissipative dynamics. 
This unitary evolution mimicking open-system behavior is a hallmark of the TFD approach and has been explored in both quantum simulation and holographic contexts \cite{qian2024quantum, Jafferis2022}.

\subsection{Temperature Dependence of the Transverse Precession}

\begin{figure}[H]
    \centering
    \includegraphics[width=\linewidth]{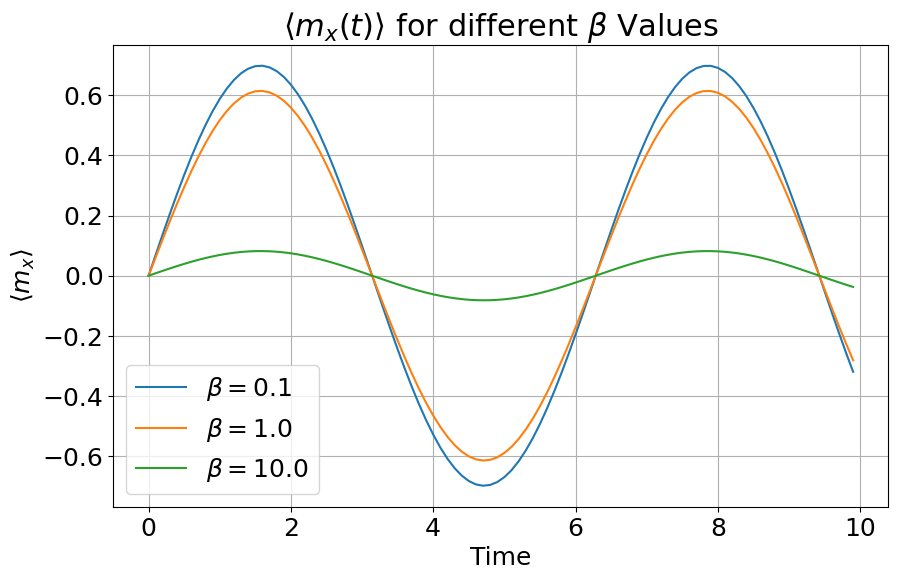}
    \caption{Time evolution of the transverse magnetization $\langle M_x(t) \rangle$ for different inverse temperatures $\beta$. The amplitude of oscillation decreases with increasing temperature, consistent with the TFD thermal damping factor $\cos\theta$.}
    \label{fig:mx_different_betas}
\end{figure}

Figure \ref{fig:mx_different_betas} illustrates the time evolution of the transverse magnetization $\langle M_x(t) \rangle$ for several values of the inverse temperature $\beta$. As $\beta$ increases (lower temperature), the amplitude of the oscillations approaches the maximum value of 1, reflecting fully coherent precession. At higher temperatures (smaller $\beta$), the amplitude decreases due to the thermal damping factor $\cos\theta$, demonstrating the expected temperature dependence of the TFD dynamics.

Therefore, this behavior quantitatively confirms that the proposed circuit correctly encodes thermal effects through the Bogoliubov transformation. 
Analogous temperature-dependent suppression of coherent dynamics has been observed in recent quantum simulations employing TFD-based methodologies \cite{Zhu2020, Wu2019}, reinforcing the validity of our approach.

In summary, the results demonstrate that the gate-based implementation reproduces both static and dynamical thermal properties with high accuracy, providing a scalable and hardware-compatible framework for finite-temperature quantum simulations. Thus, the main contribution of this work is the formulation of a deterministic and platform-agnostic quantum algorithm for the preparation and real-time evolution of TFD sttes, in which thermal effects are incorporated at the circuit level through a controlled Bogoliubov transformation. 

Unlike variational or sampling-based approaches \cite{Wu2019,lee2022variational}, the present protocol does not require classical optimization loops or ancillary thermal reservoirs, allowing for an exact encoding of thermal statistics within a unitary framework. This feature makes the method particularly suitable as a benchmark for NISQ devices, where circuit depth and noise resilience are critical constraints.

\section{Conclusion}

In this work, we presented and implemented a gate-based quantum algorithm to simulate the TFD of a thermal qubit. The proposed methodology is resource-efficient, relying exclusively on single-qubit rotations and nearest-neighbor CNOT gates, with circuit depth scaling linearly with system size, making it suitable for Noisy Intermediate-Scale Quantum (NISQ) devices.

The validation of the algorithm was performed on two fronts: the preparation of the thermal vacuum state and the simulation of its time evolution. The results for the magnetization $\langle M_z \rangle$ as a function of the inverse temperature $\beta$ agreed with the analytical prediction $M(\beta)=\tanh(\beta\omega/2)$ with high precision, confirming the correct encoding of the Gibbs ensemble. Furthermore, in the dynamical regime, the simulated spin precession exhibits the characteristic temperature-dependent suppression of coherent oscillations, which emerges naturally from the TFD formalism as a consequence of entanglement between the physical and auxiliary degrees of freedom. This provides a clear operational interpretation of thermal effects within a purely unitary framework.  In contrast to variational or sampling-based approaches, the present protocol directly encodes thermal statistics directly at the quantum-circuit level, avoiding classical optimization loops and enabling reproducible simulations with minimal computational costs.

The success of this implementation reinforces TFD as a powerful tool for the simulation of finite-temperature quantum systems on quantum hardware. Future work can explore the application of this technique to multi-body systems, such as Ising or Heisenberg models, and to the study of more complex phenomena, such as dissipative phase transitions and quantum thermodynamics. In this sense, the present results establish a minimal yet rigorous foundation for future studies of finite-temperature dynamics, open quantum systems, and thermodynamic processes on quantum computing platforms.

\begin{acknowledgments}

C. Cruz {,} {and Anna Beatriz M. de Souza} thank the Fundação de Amparo à Pesquisa do Estado da Bahia - FAPESB for its financial support (grant numbers APP0041/2023 {,} PPP0006/2024 {BOL2154/2025} and BOL2809/2025). 
G.X.A. Petronilo thank the Fundação de Amparo e Desenvolvimento da Pesquisa (grant number 5599 SAP. 4600680836).
This work has been partially funded by the projects Master's and PHD in Quantum Technologies QuIIN FCRH - QIN-FCRH-2025-5-1-1 and FCRH Extension CPQ Cybersecurity, both supported by Quantum Industrial Innovation (QuIIN) - EMBRAPII CIMATEC Competence Center in Quantum Technologies, with financial resources from the PPI IoT/Manufatura 4.0 of the MCTI grant number 053/2023, signed with EMBRAPII.

\end{acknowledgments}

\bibliography{bib.bib}
\bibliographystyle{apsrev4-1}

\end{document}